\documentclass[aip,reprint,showpacs,superscriptaddress,groupedaddress,balancelastpage,floatfix]{revtex4-2}  
\usepackage[utf8]{inputenc}
\usepackage{graphicx,dcolumn,bm,amssymb,amsmath,amsfonts,xcolor,multirow}
\usepackage{subfigure}
\usepackage[
pdfstartview=FitBV,
bookmarks=true,
bookmarksopen=true,
colorlinks =true,
linkbordercolor=blue,
linkcolor = blue,
citecolor = blue,
urlcolor=blue]{hyperref}
\usepackage[capitalize]{cleveref} 

\crefname{section}{Sec.}{Sections}
\crefname{table}{TABLE}{TABLEs.}
\crefname{figure}{FIG.}{FIGs.}

\begin{document}
	
	\title{Ion-atom-atom three-body recombination: from the cold to the thermal regime}
	\author{Marjan Mirahmadi}
	\email{mirahmadi@fhi-berlin.mpg.de}
	\affiliation{Fritz-Haber-Institut der Max-Planck-Gesellschaft, Faradayweg 4-6, D-14195 Berlin, Germany}
	\author{Jes\'{u}s P\'{e}rez-R\'{i}os}
	\email{jesus.perezrios@stonybrook.edu}
	\affiliation{ Department of Physics and Astronomy, Stony Brook University, Stony Brook, NY 11794, USA}
	\affiliation{Institute for Advanced Computational Science, Stony Brook University, Stony Brook, NY 11794, USA}
	
	\begin{abstract}
		We present a study on ion-atom-atom reaction A+A+B$^+$ in a wide range of systems and collision energies ranging from 100~$\mu$K to 10$^5$~K, analyzing the two possible products: molecules and molecular ions. The dynamics is performed via a direct three-body formalism based on a classical trajectory method in hyperspherical coordinates developed in [J. Chem. Phys. \textbf{140}, 044307 (2014)]. Our chief finding is that the dissociation energy of the molecular ion product acts as a threshold energy separating the low and high energy regimes. In the low energy regime, the long-range tail of the three-body potential dictates the fate of the reaction and the main reaction product. On the contrary, in the high energy regime, the short-range of atom-atom and atom-ion interaction potential dominates the dynamics, enhancing molecular formation for the low energy regime.

	\end{abstract}
	\maketitle
	
	\section{Introduction}\label{intro}	
	Three-body recombination, also known as ternary association, is a termolecular reaction leading to the formation of a bound state between two of the colliding particles, i.e., A+A+A$\rightarrow$A$_2$+A. Three-body recombination processes play a vital role in many areas of physics and chemistry, such as atomic and molecular processes in the ultracold regime,~\cite{Esry1999,Weiner1999,Bedaque2000,Suno2003,Weber2003,Schmidt2020,Greene2017,Koehler2006,Blume2012,Perez-Rios2015,Krukow2016,Mohammadi2021} chemical physics,~\cite{Brahms2008,Suno2009,Brahms2010,Brahms2011,Wang2011,Tariq2013,Quiros2017,Mirahmadi2021,Mirahmadi2021a} cold chemistry,~\cite{Perez-Rios2021,Perez-Rios2020} plasma physics,~\cite{Krsti2003,Cretu2022,Fletcher2007} astrophysics,~\cite{Palla1983,Flower2007,Turk2011,Forrey2013} and atmospheric physics.~\cite{Charlo2004,Luther2005,Kaufmann2006,Mirahmadi2022}
	
	
	In particular, ion-atom-atom three-body recombination processes have received much attention thanks to the recent developments in producing hybrid ion-atom systems. In the cold regime, this process (for high enough atomic densities) is the primary ion loss mechanism,~\cite{Weckesser2021,Krukow2016a,Harter2012} leading to newly formed charged products.~\cite{Perez-Rios2015,Krukow2016} Furthermore, this few-body scenario gives insight into the problem of charged impurities in an ultracold atomic gas,~\cite{Perez-Rios2021,Hirzler2020} relevant to many-body physics. Ion-atom-atom three-body recombination reactions involving rare gases are of fundamental interest in radiation physics,~\cite{Jones1980,Neves2007,Neves2010,Papanyan1995} or in the case of hydrogen and deuterium, in plasma physics.~\cite{Cretu2022,Krsti2003}
	In all the mentioned areas, with the exception of the plasma physics, the reaction occurs at temperatures $\lesssim 1$~K. As a result, most theoretical efforts have been focused on the low collision energy regime. Therefore, a comprehensive and general study of ion-atom-atom three-body processes in a wide range of collision energies is still lacking.

	

	Herein, we investigate the ion-atom-atom direct three-body reaction A+A+B$^+$, based on a classical trajectory method in hyperspherical coordinates. During this process, two different products might form: molecular ions, AB$^+$, and neutral molecules, A$_2$, from A+A+B$^+ \rightarrow$ A+AB$^+$ and A+A+B$^+ \rightarrow$ A$_2$+B$^+$ reactions, respectively. We aim to study both reaction products by comparing their formation rates based on the strengths of the long-range two-body interactions $-C_6/r^6$ (atom-atom) and $-C_4/r^4$ (ion-atom). 
	
	To this end, we introduce an effective (hyper-) radial potential in hyperspherical coordinates and find the power-dependence of this potential over a wide range of $C_6$ and $C_4$ values. Using this potential, we are able to confirm the previously derived threshold law for ion-neutral-neutral three-body recombination~\cite{Perez-Rios2015,Perez-Rios2018} at low temperatures and establish the range for its validity. Moreover, we find new and intriguing scenarios in which the branching ratio of the product states after three-body recombination deviates from the expected threshold law in the cold regime. 
	
	This paper is organized as follows: In \cref{sec1}, we introduce the Hamiltonian and explain the method. In \cref{sec2} an effective long-range radial potential has been derived to characterize the tree-body collision based on its power-dependence. Using these findings, a classical threshold law is established in \cref{sec3}. In  \cref{sec4}, we investigate the formation probabilities and recombination rates for different products through several examples of three-body reactions. Finally, \cref{sec:conclusion} provides a summary and outlines the prospects for future applications of the present work. 
	
	\section{A classical trajectory method in hyperspherical coordinates}\label{sec1}
	The dynamics of a system consisting of three particles with masses $m_i$ ($i = 1,2,3$) interacting via the potential $	V(\vec{r}_1,\vec{r}_2,\vec{r}_3)$ is governed by the Hamiltonian
	\begin{equation}\label{eq:cartesianH}
		H = \frac{\vec{p}_1^{~2}}{2m_1} + \frac{\vec{p}_2^{~2}}{2m_2} + \frac{\vec{p}_3^{~2}}{2m_3} + V(\vec{r}_1,\vec{r}_2,\vec{r}_3) ~,
	\end{equation}
	with  $\vec{r}_i$ and $\vec{p}_i$ being the position and momentum vectors of the $i$-th particle, respectively. 
	Throughout the present work we make use of the pairwise additive approximation which states that the total potential of a $N$-body system is the sum of all two-body interactions in the system. In particular, we introduce the pairwise potentials $U(r_{ij})$ for neutral-neutral interactions and $\tilde{U}(r_{ij})$ for charged-neutral interactions. As a result, the interaction potential in \cref{eq:cartesianH} read as
	\begin{equation}\label{eq:add_pot}
		V(\vec{r}_1,\vec{r}_2,\vec{r}_3) = U(r_{12}) +  \tilde{U}(r_{23}) +  \tilde{U}(r_{31}) ~,
	\end{equation}
	where $r_{ij} = |\vec{r}_j - \vec{r}_i|$.
	
\begin{figure*}[t]
		\begin{center}
			\includegraphics[scale=0.52]{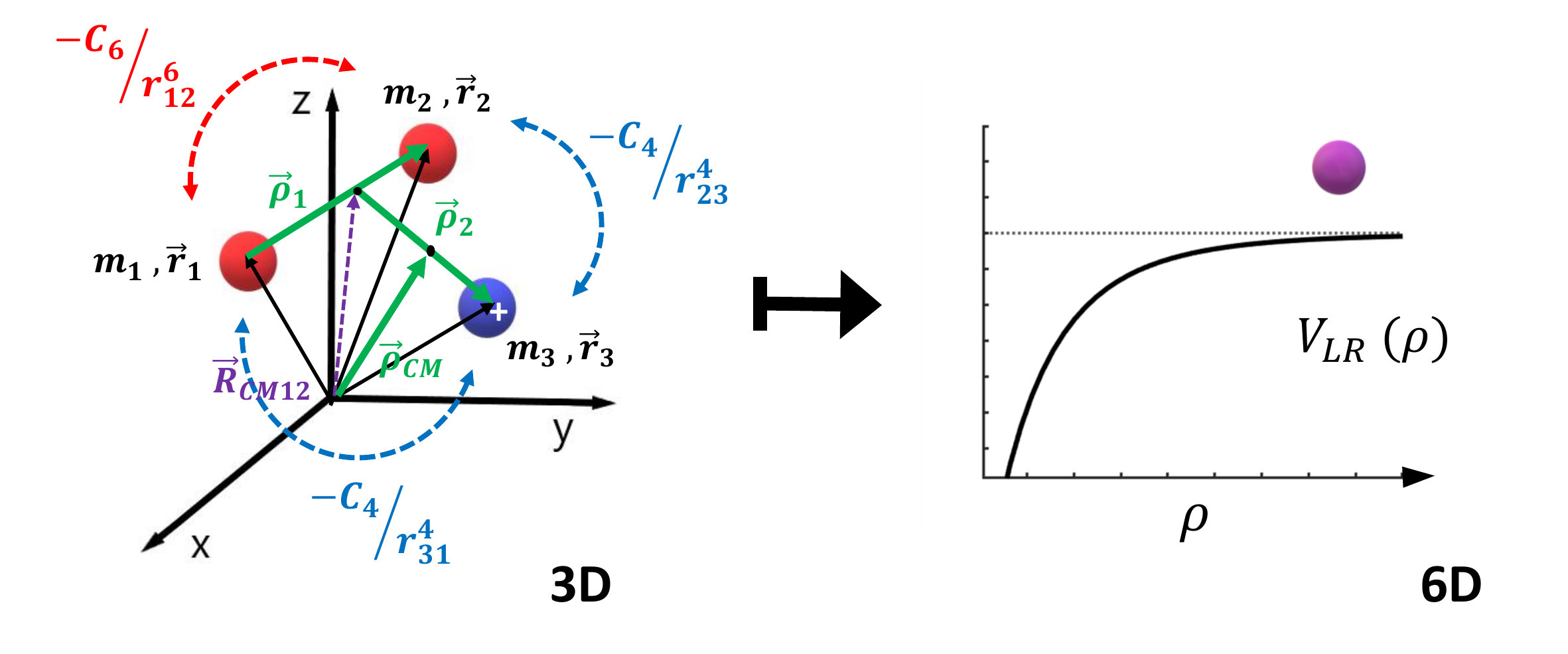}
			\caption{A schematic illustration of the long-range two-body interactions between three particles in 3D space and its counterpart, $V_{LR}(\rho)$, for a single particle in the 6D space. Jacobi coordinates for the three-body problem are shown as green vectors. Black arrows indicate the position of the three particles in Cartesian coordinates and the purple arrow indicates the two-body center-of-mass vector $\vec{R}_{CM12}$.}	
			\label{fig:jacobi}	
		\end{center}
	\end{figure*}

	It is convenient to study the three-body problem in Jacobi coordinates~\cite{Pollard1976,suzuki1998} related to the position vectors in Cartesian coordinates by the relations	
	\begin{align}\label{eq:jacobitrans}
		\vec{\rho}_1 &= \vec{r}_2 - \vec{r}_1 ~, \nonumber \\
		\vec{\rho}_2 &= \vec{r}_3 -\vec{R}_{CM12} ~, \nonumber \\
		\vec{\rho}_{CM} &= \frac{m_1\vec{r}_1 + m_2\vec{r}_2 + m_3\vec{r}_3}{M} ~,
	\end{align}
	where  $M = m_1 + m_2 + m_3$ is the total mass and $ \vec{R}_{CM12} = (m_1\vec{r}_1 + m_2\vec{r}_2)/(m_1+m_2)$  and $\vec{\rho}_{CM}$ are the center-of-mass vectors of the two-body and three-body systems, respectively. 
	The Jacobi vectors are illustrated as the green vectors in \cref{fig:jacobi} .
	Due to conservation of the total linear momentum ($\vec{\rho}_{CM}$ is a cyclic coordinate), we can omit the degrees of freedom of the center of mass. Thus, the Hamiltonian~\eqref{eq:cartesianH} will be transformed to
	\begin{equation}\label{eq:jacobiH}
		H = \frac{\vec{P}_1^2}{2\mu_{12}} + \frac{\vec{P}_2^2}{2\mu_{3,12}} +  V(\vec{\rho}_1,\vec{\rho}_2) ~,
	\end{equation} 
	with reduced masses $\mu_{12}=m_1m_2/(m_1 + m_2)$ and $\mu_{3,12}=m_3(m_1+m_2)/M$. $\vec{P}_1$ and $\vec{P}_2$ indicate the conjugated momenta of the Jacobi vectors $\vec{\rho}_1$ and $\vec{\rho}_2$, respectively. It is worth mentioning that the relations given by \cref{eq:jacobitrans} indicate a canonical transformation, and consequently, the Hamilton's equations of motion are invariant under the transformation to Jacobi coordinates. 
	
	
	\subsection{Scattering problem in hyperspherical coordinates} 
	It is well-known that an N-body collision in a three-dimensional (3D) space can be mapped onto a scattering problem of a single particle with a definite momentum moving towards a scattering center in a ($N-3$)-dimensional space. In particular, the independent relative coordinates of the three-body system, associated with the Hamiltonian~\eqref{eq:jacobiH} in the 3D space, are mapped onto the degrees of freedom of a single particle moving towards a scattering center in a six-dimensional (6D) space. We choose a 6D space parametrized by hyperspherical coordinates consisting of a hyperradius $R$, and five hyperangles $\alpha_j$ (with $j = 1,2,3,4,5$), where $0\leq\alpha_1<2\pi$ and $0\leq\alpha_{j>1}\leq\pi$.~\cite{Lin1995,Avery2012,Perez-Rios2014} 	
	The volume element in this coordinate system is given by
	\begin{align}\label{eq:hsvol}
		d\tau &= R^5d Rd\Omega \nonumber \\
		&= R^5d R\prod_{j=1}^{5}\sin^{j-1}(\alpha_j)d\alpha_j ~.
	\end{align} 
	The position and momentum vectors in this space can be constructed from the Jacobi vectors and their conjugated momenta as~\cite{Perez-Rios2014,Perez-Rios2020}
	\begin{equation}\label{eq:rho6D}
		\vec{\rho} = \begin{pmatrix} \vec{\rho}_1 \\ \vec{\rho}_2 \end{pmatrix}
	\end{equation}
	and 
	\begin{equation}\label{eq:P6D}
		\vec{P} = \begin{pmatrix} \sqrt{\frac{\mu}{\mu_{12}}}\vec{P}_1 \\ \sqrt{\frac{\mu}{\mu_{3,12}}}\vec{P}_2 \end{pmatrix} ~,
	\end{equation}
	respectively. Here $\mu = \sqrt{m_1 m_2 m_3/ M}$ is the three-body reduced mass. By using \cref{eq:rho6D,eq:P6D}, the Hamiltonian in the 6D space reads as
	\begin{equation}\label{eq:6DH}
		H^\mathrm{6D} = \dfrac{\vec{P}^2}{2\mu} + V(\vec{\rho}) ~.
	\end{equation}
	
	The concept of classical cross section $\sigma$ for the scattering problem in the 3D space can be extended to the 6D space by visualizing it as an area in a five-dimensional hyperplane (embedded in the 6D space) perpendicular to the initial (6D) momentum vector $\vec{P}_0$. Thus, the impact parameter vector $\vec{b}$ in the 6D space can be defined as projection of the initial position vector $\vec{\rho}_0$ on this hyperplane. Therefore the necessary condition $\vec{b}\cdot\vec{P}_0 = 0$ is satisfied. 
	
	Note that, by treating three-body collision as a scattering problem of a single particle in a 6D space, we can define the initial conditions and the impact parameter uniquely as single entities (in the 6D space). Therefore, it is possible to characterize the outcome of a three-body process as a function of the impact parameter $\vec{b}$ and the initial momentum $\vec{P}_0$. In particular, for three-body recombination, the total cross section is given by~\cite{Perez-Rios2014,Perez-Rios2020}
	\begin{align}\label{eq:sigma}
		\sigma_\mathrm{rec}(E_c) & = \dfrac{\int \mathcal{P}(\vec{P}_0,\vec{b})  b^4 d b ~d\Omega_bd\Omega_{P_0}}{\int d\Omega_{P_0}} \nonumber \\
		&= \frac{8\pi^2}{3}\int_{0}^{b_\mathrm{max}(E_c)} \mathcal{P}(E_c,b)  b^4 d b ~,
	\end{align}
	after averaging over different orientations of $\vec{P}_0$. In \cref{eq:sigma}, $d\Omega_b$ and $d\Omega_{P_0}$ denote the differential elements of the solid hyperangle associated with vectors $\vec{b}$ and $\vec{P}_0$, respectively, where $\Omega_b = 8\pi^2/3$. The so-called opacity function $\mathcal{P}$ in \cref{eq:sigma} is the probability of a recombination event as a function of the impact parameter $b$ and collision energy $E_c$ (obtained from $E_c=P_0^2/(2\mu)$). The angular dependence of the opacity function $\mathcal{P}(\vec{P}_0,\vec{b})$, which depends on both direction and magnitude of impact parameter and initial momentum vectors, has been averaged out by means of Monte Carlo method explained further below. $b_\mathrm{max}$ represents the largest impact parameter for which three-body recombination occurs, or in other words, $\mathcal{P}(E_c,b) = 0$ for $b>b_\mathrm{max}$.
	Consequently, the energy-dependent three-body recombination rate is given by
	\begin{equation}\label{eq:k3}
		k_3(E_c) = \sqrt{\frac{2E_c}{\mu}}\sigma_\mathrm{rec}(E_c) ~.
	\end{equation}

	\subsection{Computational details}\label{subsec:comput}
	The initial orientation of vectors $\vec{P}_0$ and $\vec{b}$ in the 6D space are sampled randomly from probability distribution functions associated with the appropriate angular elements in hyperspherical coordinates (see Ref.~[\onlinecite{Perez-Rios2020}]). For the sake of simplicity and without loss of generality, we choose the $z$ axis in 3D space to be parallel to the Jacobi momentum vector $\vec{P}_2$. Note that the condition $\vec{b}\cdot\vec{P}_0 = 0$ is also implemented in the calculations.
	
	The opacity function $\mathcal{P}(E_c,b)$ for a given collision energy $E_c$ and magnitude of impact parameter $b$, is achieved by dividing the number of classical trajectories that lead to the recombination events, $n_r$, by the total number of trajectories simulated $n_t$.\cite{Perez-Rios2014} Thus,
	\begin{align}\label{eq:opacity}
		\mathcal{P}(E_c,b) \approx ~ & \frac{n_r(E_c,b)}{n_t(E_c,b)} \pm \nonumber \\
		& \frac{\sqrt{n_r(E_c,b)}}{n_t(E_c,b)}\sqrt{\frac{n_t(E_c,b)-n_r(E_c,b)}{n_t(E_c,b)}} ~,
	\end{align}
	where the second term in \cref{eq:opacity} is the statistical error owing the inherent stochastic nature of the Monte Carlo technique. For the results reported in this work, for each initial pair of $(E_c,b)$, the number of total trajectories varies between $n_t=3\times10^3$ and $n_t=10^5$ to keep the relative error in calculated $k_3(E_c)$ rate coefficients, below 5$\%$.
	
	For the results presented here, the Hamilton's equations have been solved by using the ``ode113'' of Matlab ODE suite. This is a variable-step/variable-order predictor–corrector (PECE of orders 1 to 13) implementation of the  Adams-Bashforth-Moulton methods.\cite{Shampine1997} The acceptable error for each time-step has been determined by absolute and relative tolerances equal to $10^{-15}$ and $10^{-13}$, respectively. The total energy is conserved during collisions to at least four significant digits and the magnitude of the total angular momentum vector, $J = |\vec{\rho}_1 \times \vec{P}_1 + \vec{\rho}_2 \times \vec{P}_2|$, is conserved to at least six significant digits. The initial magnitude of hyperradius, $|\vec{\rho}_0|$, is generated randomly from the interval $[R_0-\delta R, R_0+\delta R]~a_0$ centered around a suitable $R_0$ which fulfils the condition for three particles to be initially in an uniform rectilinear state of motion. Here, $a_0$ is the Bohr radius ($\approx 5.29 \times 10^{-11} \mathrm{m}$).
	
	\section{Long-range (hyper-) radial potential}\label{sec2} 
	\begin{figure*}
		\begin{center}
			\includegraphics[scale=0.45]{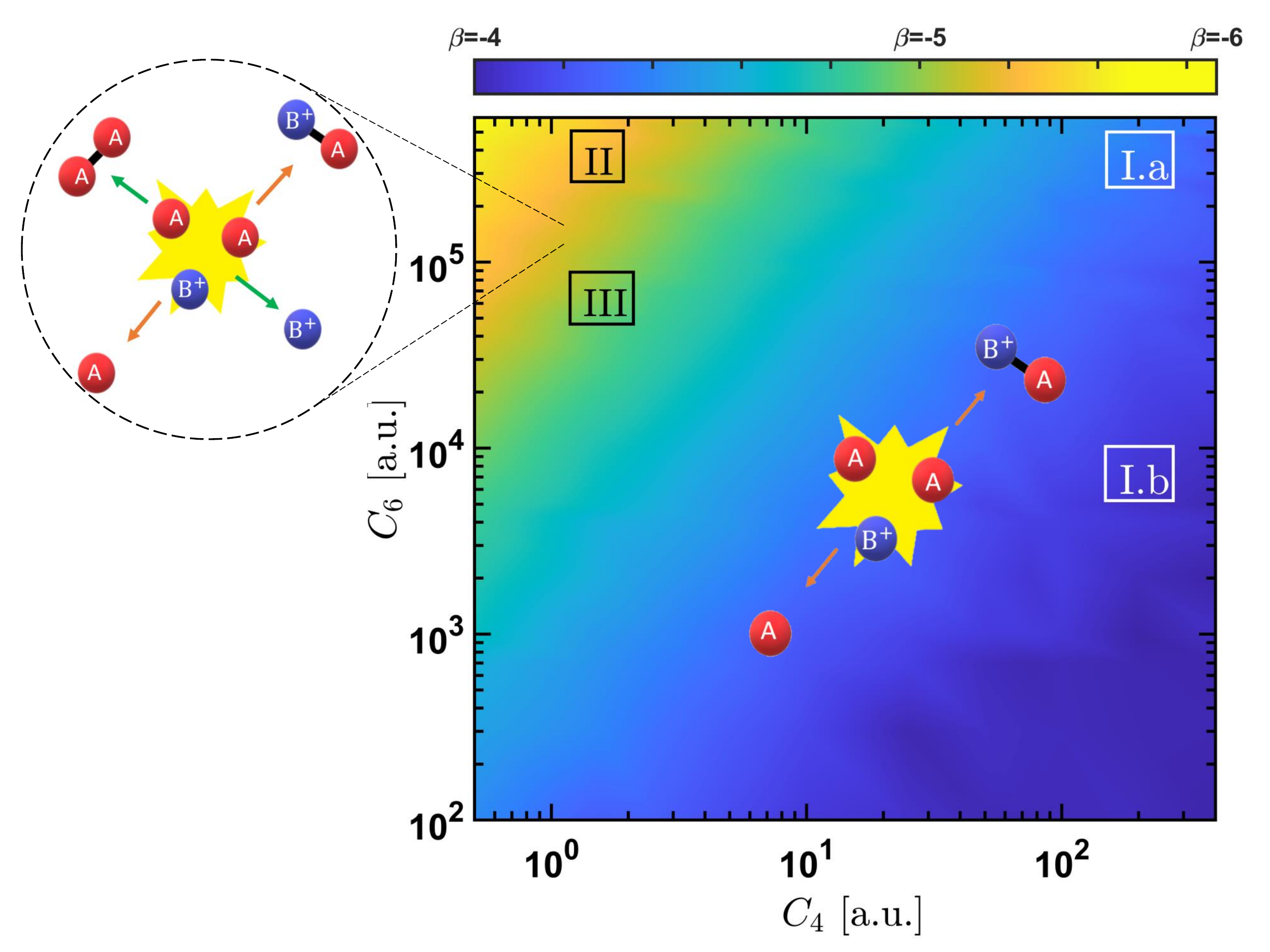}
			\caption{\label{fig:Pdiag} Heat map of visualizing the parameter $\beta$ as a function of long-range two-body interaction coefficients $C_4$ and $C_6$ in the log-log scale. Letters indicate the examples chosen from different regimes ($\beta=-4, -5,$ and $-6$). The schematic illustrations display the dominant reactions at low collision energies.}
		\end{center}
	\end{figure*}
	It is possible to characterize the A+A+B$^+$ three-body recombination reaction and its products at low temperatures, based on the long-range behavior of the two-body potentials, i.e.,  $U(r_{12})\rightarrow -C_6/r_{12}^6$ for A$_2$ and $\tilde{U}(r_{23})\rightarrow -C_4/r_{23}^4$ and $\tilde{U}(r_{31})\rightarrow -C_4/r_{31}^4$ for AB$^+$. To this end, we find the corresponding long-range potential in the 6D space relevant for the classical trajectory method explained in the previous section.       
	Hence, the effective long-range potential in hyperspherical coordinates can be obtained from the following relation (see \cref{fig:jacobi}),
	\begin{equation}\label{eq:vlr}
		V_{LR}(\vec{\rho}) = -\frac{C_6}{r_{12}^6}-\frac{C_4}{r_{23}^4}-\frac{C_4}{r_{31}^4}~,
	\end{equation}
	where $C_6=C_6^\mathrm{A_2}$ is the van der Waals dispersion coefficient and $C_4=C_4^\mathrm{AB^+}$ is half of the atom (A) polarizability (in atomic units). 
	
	Noting \cref{eq:jacobitrans,eq:rho6D}, potential $V_{LR}(\vec{\rho})$ depends on the magnitude of the 6D position vector, $\rho = |\vec{\rho}|$, as well as the hyperangles  $(\alpha_1,\alpha_2,\alpha_3,\alpha_4,\alpha_5)$ associated with it. Thus, to find the radial dependence of this potential, labelled as $V_{LR}(\rho)$ in the schematic illustration in \cref{fig:jacobi}, we solve \cref{eq:vlr} for randomly sampled hyperangles with appropriate weights (given in \cref{eq:hsvol}), ensuring a uniform sampling of the configuration space (for more details see Refs.~[\onlinecite{Mirahmadi2021a,Wang2022}]). Considering $C_4$ and $C_6$ constants, the (hyper-) radial potential reads as,
	
	\begin{equation}
		V_{LR} (\rho)=	-C_\mathrm{eff} \rho^{\beta}~. 
	\end{equation}
	Consequently, the power $\beta$ can be considered as a function $\beta(C_6,C_4)$.

	
	\Cref{fig:Pdiag} displays the parameter $\beta$ as a function of $C_6\in [10^2,6\times10^5]$ and  $C_4\in[0.5,400]$, in atomic units.  In this figur,e we identify three main regimes, associated with $\beta\approx-6$ (yellow color), $\beta\approx-4$ (blue color), and an intermediate regime $\beta\approx-5$ (greenish yellow color). Different values of $\beta$ translate into the preponderance of a given reaction product, as shown below. In particular, $\beta\approx-4$ represents a typical scenario in which the charged-neutral interaction dominates the course of the reaction, leading mainly to the formation of ions, as sketched in \cref{fig:Pdiag}. On the contrary, $\beta\approx-6$ means that the neutral-neutral interaction is the most significant interaction, which translates into a larger production of neutral molecules. 
	
	Surprisingly enough, there is a last scenario in which both neutral-neutral and neutral-charged interaction have a considerable contribution leading to $\beta\approx-5$. In such a case, the three-body recombination should lead to a similar amount of neutral molecules to molecular ions. However, this is an unexpected scenario since the long-range two-body potentials are proportional to $r_{ij}^{-4}$ and $r_{ij}^{-6}$ for charged-neutral and neutral-neutral interactions, respectively, but the hyper-radial potential has the power-dependence $\rho^{-5}$. 
	
	It is worth mentioning that the coefficients $C_6$ and $C_4$ in most ion-atom-atom reactions are associated with $\beta\approx-4$.

	\section{Generalized classical threshold law}\label{sec3}
	The general trend of the three-body recombination rate as a function of the collision energy ($E_c$) fulfills a threshold law in the low-energy regime. In particular, using the fact that the long-range tail of the potential dominates the recombination rate at low energies, we can derive a classical threshold law associated with the quantum $s$-wave scattering, i.e., zero quantum angular momentum. In classical scattering, one may define the maximum impact parameter, $b_\mathrm{max}$, as the distance at which the collision energy is comparable to the strength of the interaction potential, i.e., $E_c = C_6r_{12}^{-6} + C_4r_{23}^{-4} + C_4r_{31}^{-4} $, in 3D space, or equivalently $E_c =  C_\mathrm{eff}\rho^\beta$, in 6D space. Note that the coefficient $C_\mathrm{eff}$ can be obtained for different values of $\beta$ (for more details, see Ref.~[\onlinecite{Mirahmadi2021a}]), however, here we are only interested in the power-law dependence of the $k_3(E_c)$. Therefore, we derive the following relation for $b_\mathrm{max}$,  
	\begin{align}\label{eq:bmax}
		b_\mathrm{max} \propto E_c^{1/\beta}.
	\end{align}
	
	The geometric cross section is obtained by setting $\mathcal{P}(E_c,b) = 1$ for $b\leq b_\mathrm{max}$ (also known as rigid-sphere model) in \cref{eq:sigma}. Thus, upon substituting \cref{eq:bmax} into \cref{eq:sigma}, we find the energy-dependence of the geometric cross section as, 
	\begin{align}\label{eq:sigmaLang}
		\sigma_\mathrm{rec}(E_c) & = \frac{8\pi^2}{3}\int_{0}^{b_\mathrm{max}(E_c)} b^4 d b  \propto E_c^{5/\beta} ~.
	\end{align}
	Employing \cref{eq:k3}, the three-body recombination rate can be calculated as a function of collision energy, 
	
	\begin{equation}\label{eq:k3Lang}
		k_3(E_c) \propto E_c^{(10+\beta)/(2\beta)} ~.
	\end{equation}
	
	Setting $\beta=-4$, \cref{eq:sigmaLang,eq:k3Lang} lead to  $\sigma_\mathrm{rec}(E_c)\propto E_c^{-5/4}$ and $k_3(E_c) \propto E_c^{-3/4}$. This result verifies the threshold law given in Refs.~[\onlinecite{Perez-Rios2015,Perez-Rios2018}], which has been obtained under the assumption that only ion-atom interaction dictates the outcome of the three-body recombination. This is in accordance with our findings displayed in \cref{fig:Pdiag} and the related discussion in \cref{sec2}. Note that the rate given by \cref{eq:k3Lang} accounts for both A$_2$ and AB$^+$ products of the three-body recombination. However, as it is discussed below, in this scenario, AB$^+$ molecules are the main reaction product. In the two other regimes, i.e., $\beta=-5$ and  $-6$, the power-law yields $k_3(E_c) = E_c^{-1/3}$ and $k_3(E_c) = E_c^{-1/2}$, respectively. 
	
	\section{Results and discussion}\label{sec4}
	The three-body recombination process A+A+B$^+$ might result in one of two different products, namely, the molecular ion, AB$^+$, and the neutral molecule, A$_2$. Molecular ions form through the reaction A+A+B$^+ \rightarrow$ A+AB$^+$, whereas neutral molecule formation follows A+A+B$^+ \rightarrow$ A$_2$+B$^+$. In this section, we investigate each reaction's importance by using the opacity function, i.e., the probability of formation of each product as a function of the collision energy, $E_c$, and the impact parameter, $b$.
	
	\subsection{Low-energy regime}\label{sub:lowE}
	We consider three different scenarios regarding the strengths of the long-range A$_2$ ($-C_6/r^6$) and AB$^+$ ($-C_4/r^4$) interactions, characterized by the parameter $\beta$ introduced in \cref{sec2}. Note that this characterization is only valid for the low-energy regime, at which the long-range interactions dictate the outcome of the three-body recombination reaction. In general, this region is assumed to correspond to the cold regime, i.e, $E_c\lesssim 1$~K.
	
	
	Here, we calculate the opacity functions for four different scenarios, labeled in \cref{fig:Pdiag}: two examples, I.a and I.b, from the regime where charged-neutral interaction is dominant ($\beta\approx-4$); the example II, for $\beta\approx-6$ where the neutral-neutral interaction is stronger; and example III for the intermediate region, i.e., $\beta\approx-5$. The corresponding $C_6$ and $C_4$ parameters are listed in \cref{tab1} and the results are shown in \cref{fig:c4dom,fig:c6dom,fig:c5dom}. The relative error due to one standard deviation error, as customary in Monte Carlo simulations (see \cref{eq:opacity}), for the lowest impact parameter $b=0$ is $\approx 1\%$ and for the maximum impact parameter, $b_\mathrm{max}$, is $\approx 5\%$.
	
	\begin{table}
		\caption{\label{tab1} Long-range coefficients of the pairwise potentials for four different regions highlighted in \cref{fig:Pdiag} for three-body recombination reactions, A+A+B$^+$. The values are given in atomic units.}
		\begin{ruledtabular}
			\begin{tabular}{cccc}
				$\beta$	& label  & $C_6$  &  $C_4$ \\
				\hline
				\multirow{2}{*}{-4}& I.a  & $6\times10^5$ & 200 \\
					& I.b  & 6640 & 200  \\
				-6	& II  & $7\times 10^4$ & $1.85$  \\
				-5	& III & $6\times10^5$ & $1.85$  \\
				\end{tabular}
		\end{ruledtabular}
	\end{table}
	
	\subsubsection{Case I: Charged-neutral-dominated processes}
	
	\begin{figure}
		\centering
		\subfigure[case I.a in \cref{fig:Pdiag}]{\includegraphics[scale=.4]{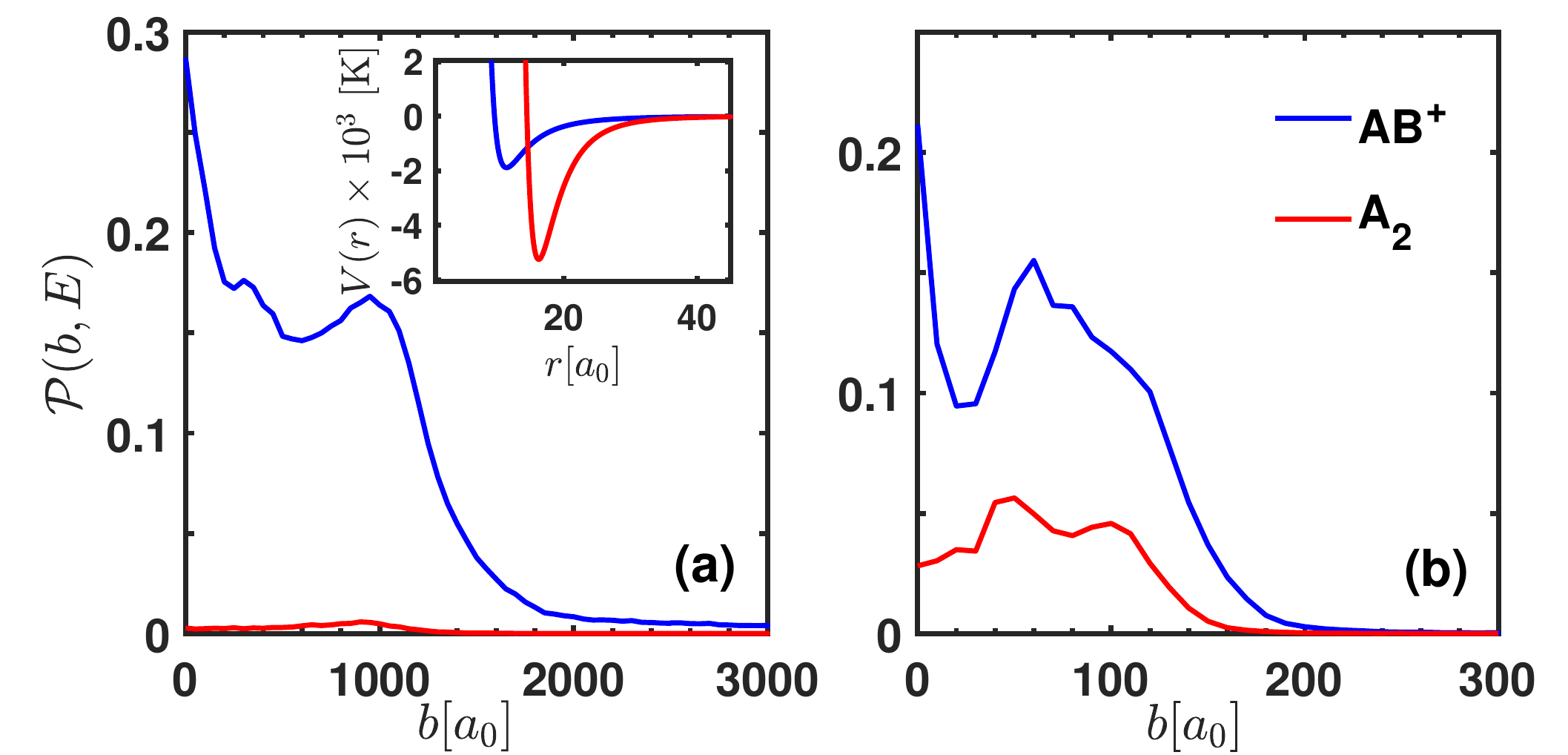}} 
		\subfigure[case I.b in \cref{fig:Pdiag}]{\includegraphics[scale=.4]{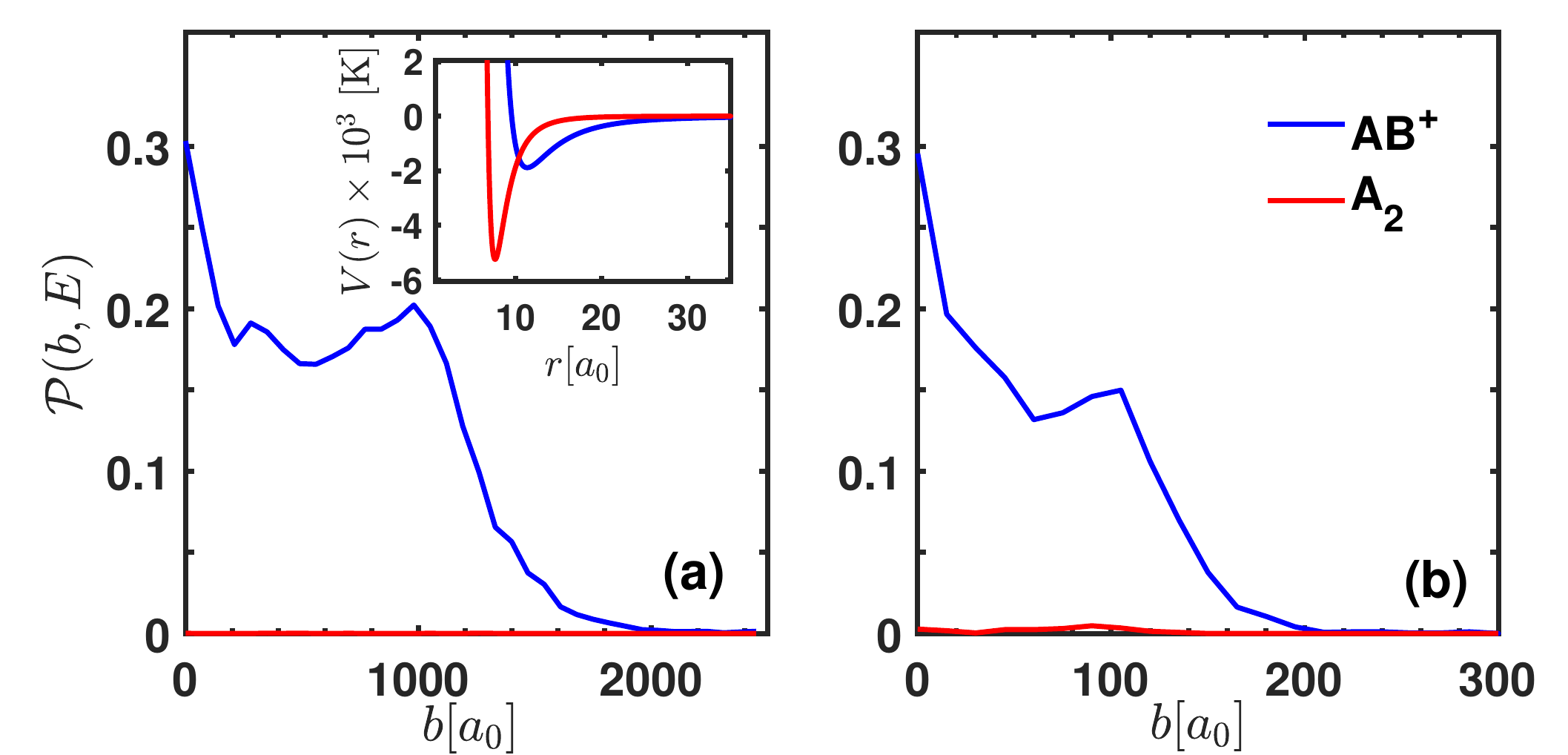}} 
		\caption{The opacity function of each reaction products for $\beta \approx -4$ at collision energies $E_c = 1$~mK (left panels) and $E_c = 10$~K (right panels). The mass of the atom is the same as $^{133}$Cs and the one for the ion corresponds to $^{87}$Rb$^+$. Pairwise potentials are shown in the inset. Here, $a_0 \approx 5.29 \times 10^{-11} \mathrm{m}$ is the Bohr radius.}
		\label{fig:c4dom}
	\end{figure}
	
	\Cref{fig:c4dom} shows the opacity functions of both products for two collision energies: 1~mK (left panel) and 10~K (right panel), for the I.a and I.b cases described above.  The figure shows that although both systems show a significant difference in the neutral-neutral interaction (the $C_6$ in I.a is approximately 100 times larger than in I.b), AB$^+$ is the main product, regardless of the collision energy. At $E_c=1$~mK, the opacity function for molecular ions ( blue curve) is the same for the two cases under consideration. However, at $E_c=10$~K, the opacity function changes from case to case. For instance, at $b=0$ and 10~K, the formation of molecular ions for the I.a case is 33$\%$ more probable than in the I.b case. The same trend, although more abrupt, is observed for the opacity associated with molecule formation. In particular, at $E_c=10$~K, I.a shows a somewhat substantial probability of formation of A$_2$ than I.b (where $\mathcal{P}$ of A$_2$ is $\approx0$) due to a larger $C_6$ value.
	
	
	
	The opacity functions for the same long-range coefficients as in examples I.a and I.b, but with different short-range interaction potentials, have been calculated, and similar results have been obtained. This confirms that the short-range region of the pairwise interaction potential does not play a role in the three-body recombination rate at the low-energy regime. 
	
	\subsubsection{Case II: Neutral-neutral-dominated processes}
	
	\begin{figure}
		\begin{center}
			\includegraphics[scale=0.45]{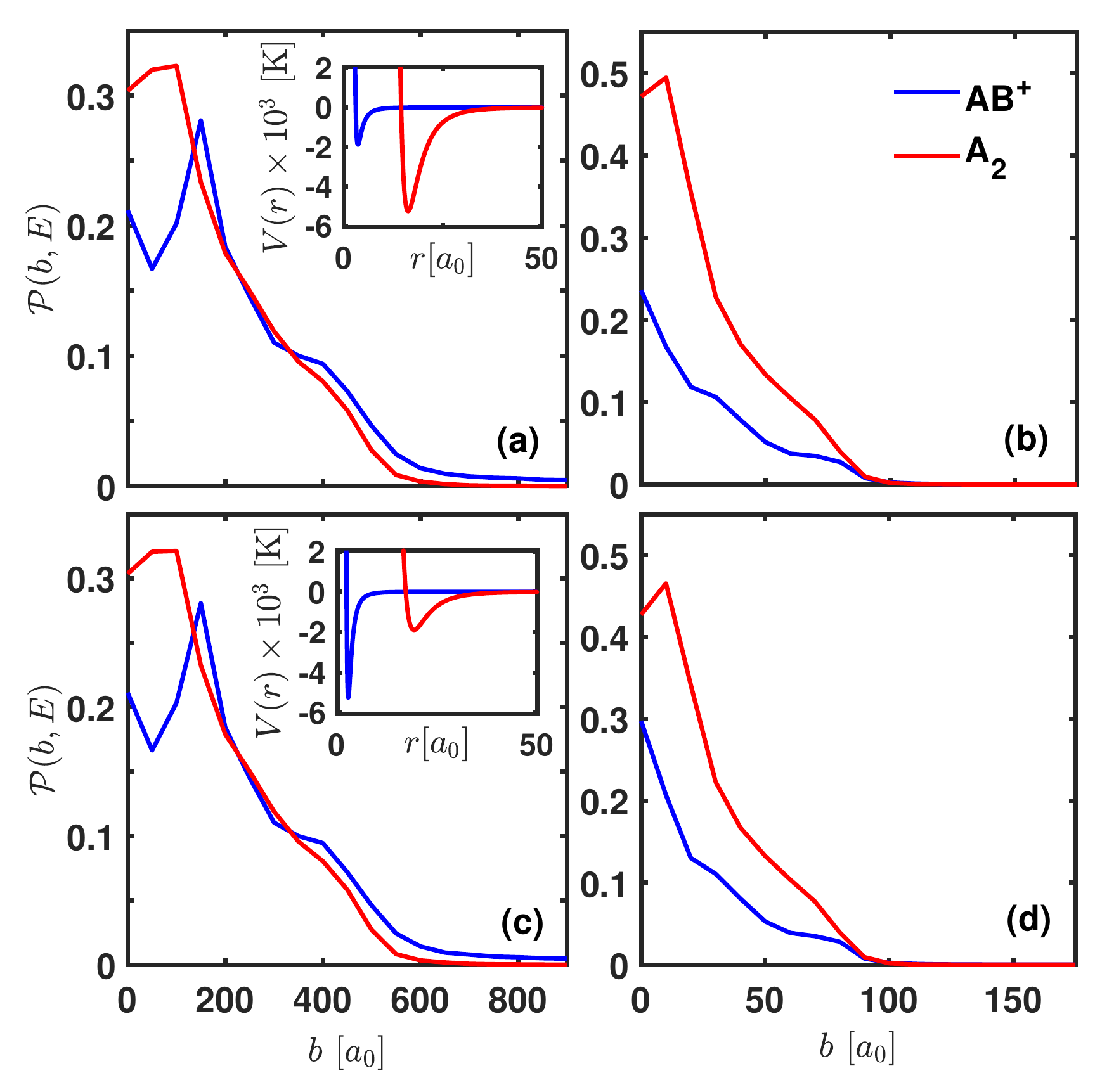}
			\caption{\label{fig:c6dom} The opacity function of each product for $\beta \approx -6$ (case II in \cref{fig:Pdiag}) and collision energies $E_c = 1$~mK (panels \text{a,c}) and $E_c = 10$~K (panels \text{b,d}).  The mass of the atom is the same as $^{133}$Cs and the one for the ion corresponds to $^{87}$Rb$^+$. Plots in each row are calculated for different cases by changing the short-range properties of the pairwise potentials shown in the inset.}
		\end{center}
	\end{figure}
	
	As it can be seen in \cref{fig:c6dom}, when $\beta = -6$ (case II in Fig.~\ref{fig:Pdiag}), there is a boost in the formation of neutral molecules regardless of the collision energy. At $E_c=1$~mK (panels a and c), A$_2$ and AB$^+$ are formed with nearly the same probability. Indeed, for small impact parameters, the production of neural molecules overcomes that of molecular ions. The ratio between the formation of neutral molecules versus molecular ions increases at $E_c=10$~K (panels b and d). Therefore, a system within $\beta = -6$ regime will show a larger molecular formation rate than in the case of $\beta = -4$.
	
	Comparing panel (a) with panel (c) and panel (b) with panel (d), we notice that for each collision energy, the opacities remain unchanged independently of the nature of the short-range neutral-neutral or charged-neutral interactions. In other words, the short-range of the potential does not affect the three-body recombination reaction rate at low energy collisions, as in the case of charged-neutral dominated processes. Finally, it is worth remarking that our results do not identify the A+A+B$^+ \rightarrow$ A$_2$+B$^+$ process as the primary reaction. However, unlike cases I.a and I.b ($\beta\approx -4$), the effect of this reaction is not negligible. 
	
	

	\subsubsection{Case III: The intermediate region}
	
	\begin{figure}
		\begin{center}
			\includegraphics[scale=0.47]{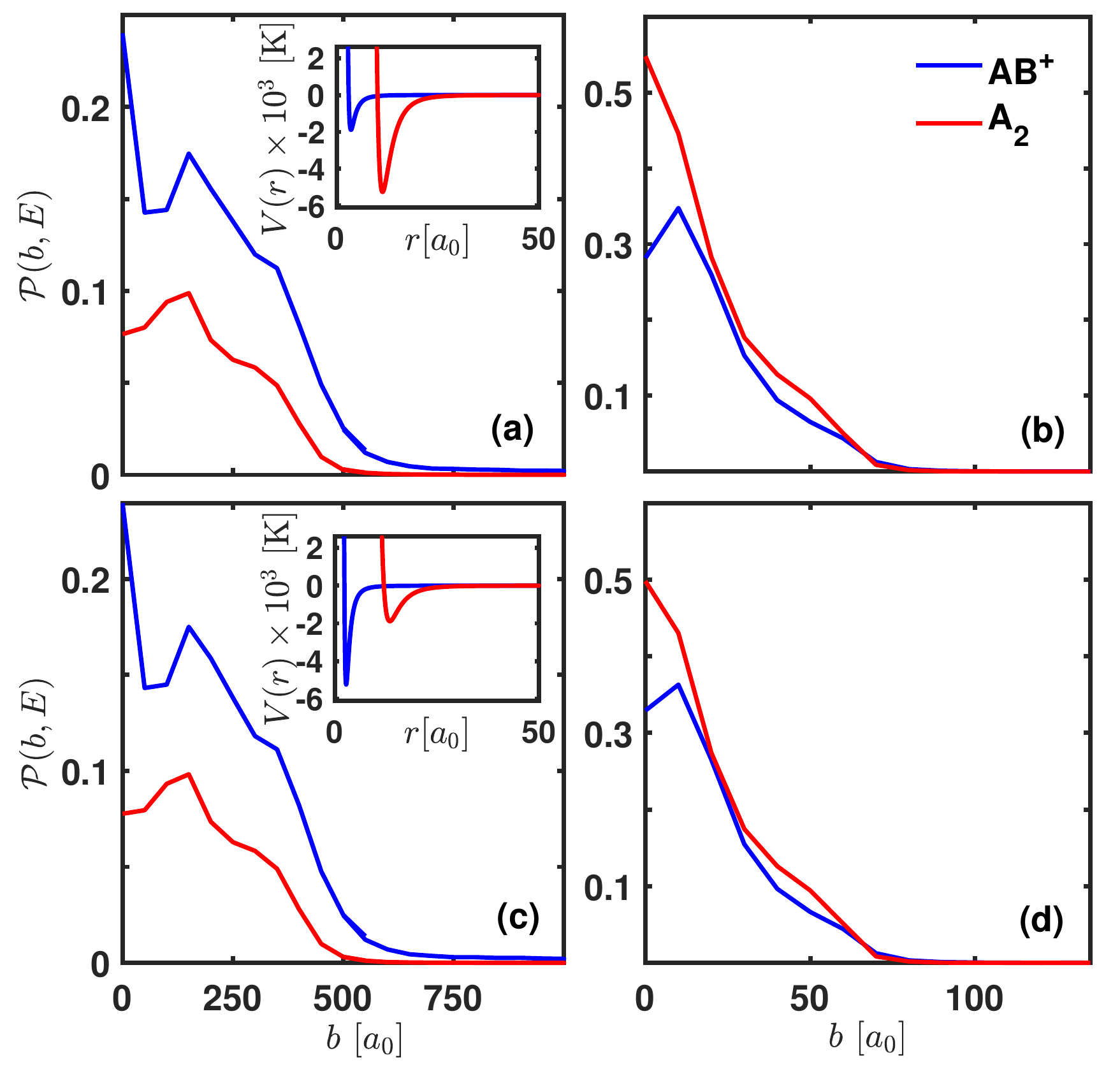}
			\caption{\label{fig:c5dom} Same as \cref{fig:c6dom} but for $\beta \approx -5$ (case III in \cref{fig:Pdiag}).}
		\end{center}
	\end{figure}
	
	The opacity functions related to the intermediate region, $\beta = -5$, for two different collision energies, are displayed in \cref{fig:c5dom}. For $E_c=1$~mK (panels (a) and (c)), we see that, even though the dominant product is AB$^+$, there is a considerable probability of formation of A$_2$. For $E_c=10$~K, the formation probabilities of neutral molecules and molecular ions are very close, except for small impact parameters ($b<20$~a$_0$), where three-body recombination is prone to molecular formation in detriment of molecular ions.  
	
Comparing \cref{fig:c4dom,fig:c6dom,fig:c5dom}, one can conclude that for systems with $\beta=-4$, the three-body recombination leads primarily to the formation of molecular ions with a negligible probability of the formation of neutral molecules. On the contrary, for systems with $\beta\approx-6$ or $\beta\approx-5$, molecular and molecular ion formation probabilities are comparable, and under certain conditions, the three-body recombination favors neutral molecule formation over molecular ions. A summary of our findings regarding the importance of different reactions in the low-energy regime has been illustrated schematically in \cref{fig:Pdiag} for the three different regions discussed above.
	
	
	
	\subsection{High-energy regime}
	In this section, we investigate three-body recombination processes at collision energies higher than previously considered. For these high energies, as we will show, the short-range region of the pairwise interaction plays a pivotal role in the reaction dynamics. Therefore, categorizing collisions based on the long-range tail of the potentials is no longer valid. 
	
\begin{figure}[b]
		\centering
		\subfigure[case I.a in \cref{fig:Pdiag}]{\includegraphics[scale=.42]{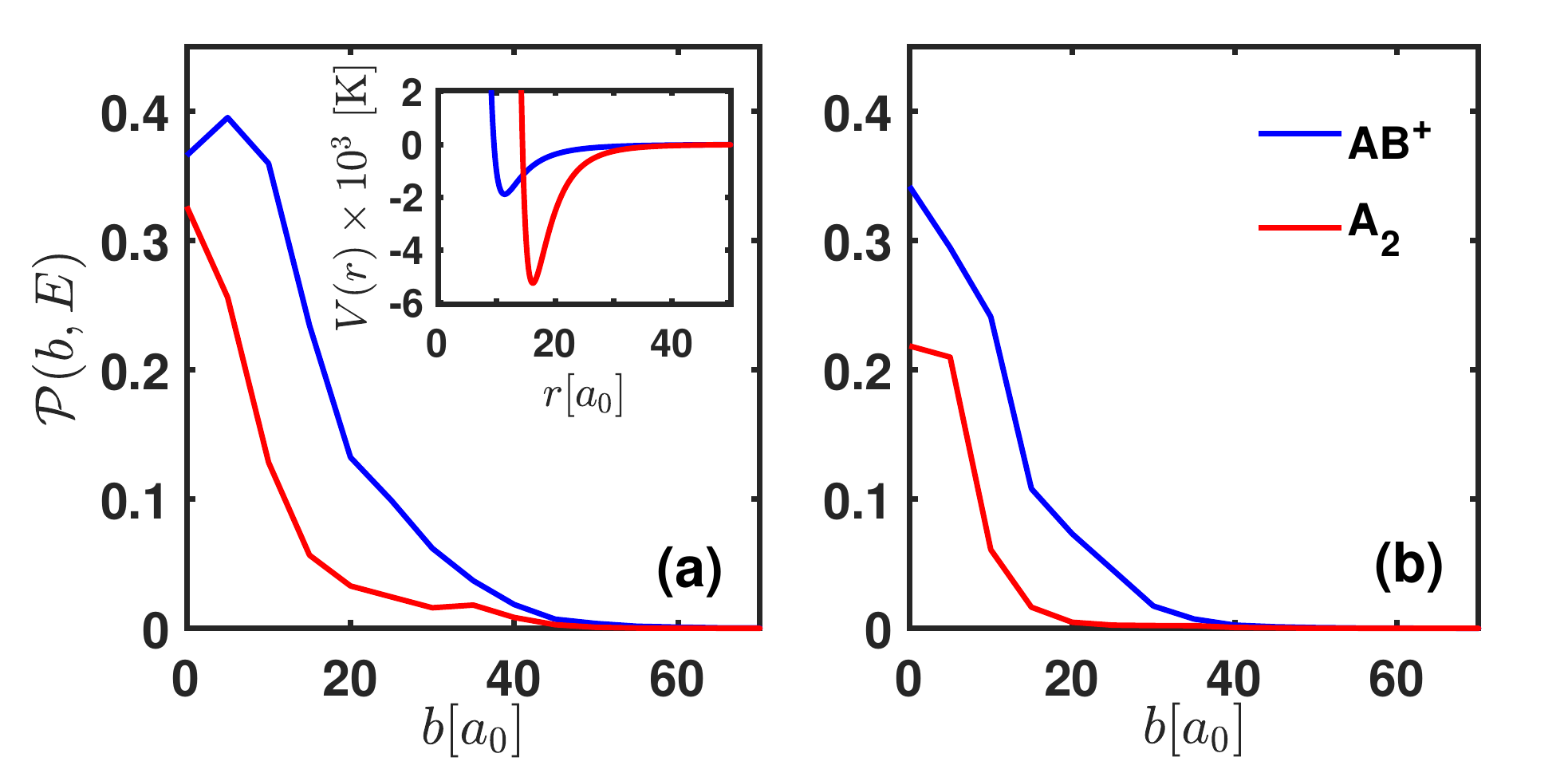}} 
		\subfigure[case I.b in \cref{fig:Pdiag}]{\includegraphics[scale=.42]{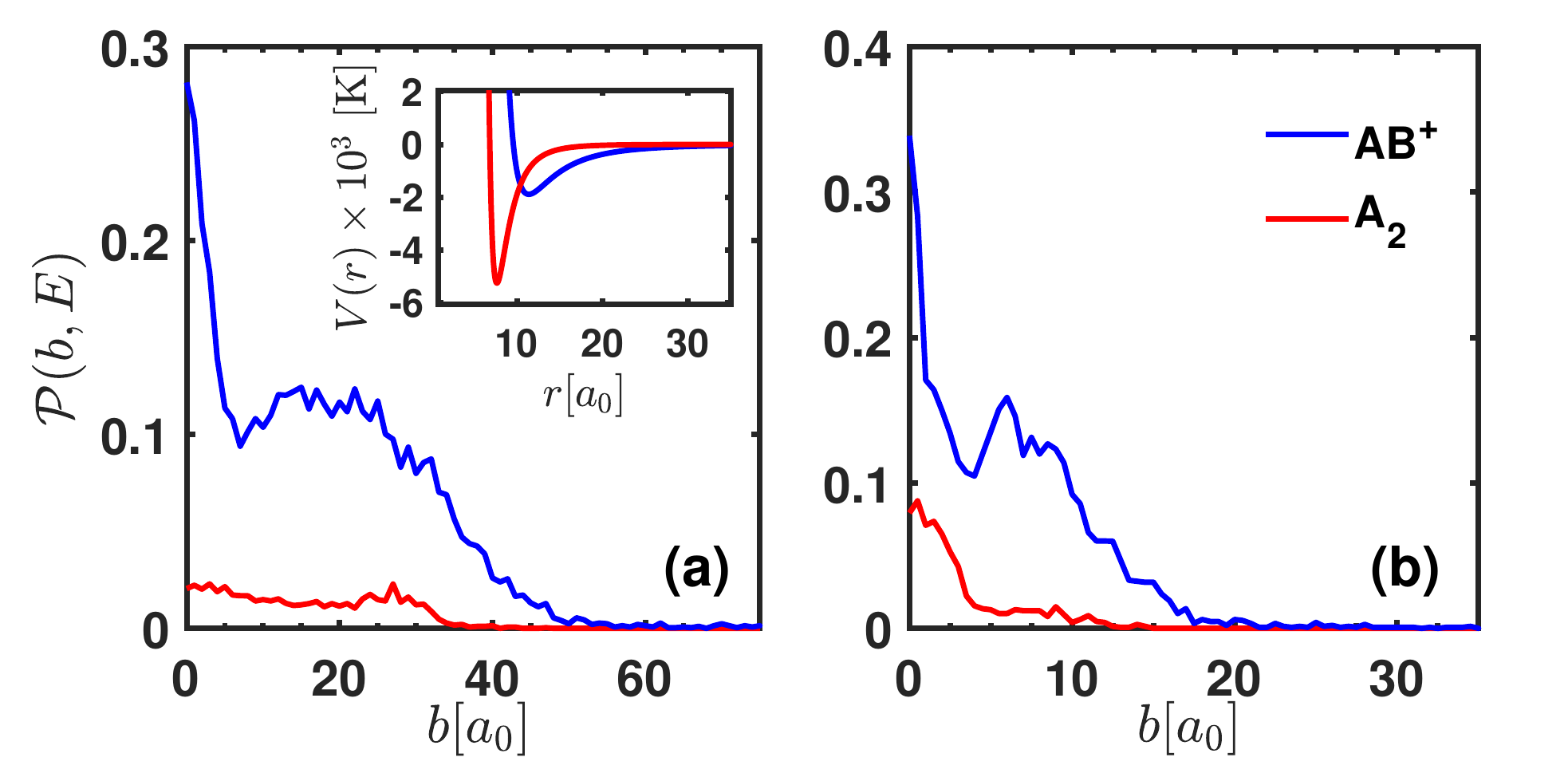}}
		\caption{Same as \cref{fig:c4domcdf} but for collision energies (left panels) $E_c=3000$~K and (right panels) $E_c=7000$~K.}
		\label{fig:c4domdfHT}
	\end{figure}

	We calculate the opacity function for two systems at two different collision energies ($E_c=3000$~K and $E_c=7000$~K). In particular, the long-range tail of the charged-neutral and neutral-neutral potentials correspond to cases I.a and I.b in Fig.~\ref{fig:Pdiag}. The charged-neutral short-range potential is the same, whereas the neutral-neutral short-range potential varies. The results are shown in \cref{fig:c4domdfHT}, where a more significant production of neutral molecules appears for the whole range of the impact parameter compared with \cref{fig:c4dom}. However, in virtue of the classical threshold law, three-body recombination should mostly lead to the formation of molecular ions since $\beta\approx-4$. Therefore, the short-range of the pairwise potential must play a major role for $E_c=3000$~K and $E_c=7000$~K. In other words, the systems under consideration enter into a new regime at high collision energies dominated by short-range physics.  
	
		To characterize the transition between low-energy to high-energy regimes, it is necessary to study the formation of the two products, A$_2$ and AB$^+$ over a wide range of collision energies $E_c$, which is the goal of the next section.

	
	

	
	\subsection{Study of representative systems}
	
In this section, we focus on four ion-atom-atom systems. Three of them representative of cold chemistry experiments in hybrid ion-atom traps, whereas the fourth is an important scenario for ion-mobility experiments. The energy-dependent three-body recombination rates are calculated via the classical trajectory method introduced in \cref{sec1} over a wide range of collision energies between $10^{-4}$~K and 10$^5$~K. Note that in these calculations, the two-body potentials are of the form $U(r) = -C_6/r^6 + C_12/r^{12}$ for the atom-atom (A-A) interaction, and $\tilde{U}(r) = -C_4/r^4 + C_8/r^8$ for the ion-atom one (A-B$^+$). 


	
	\begin{figure}
		\begin{center}
			\includegraphics[scale=0.45]{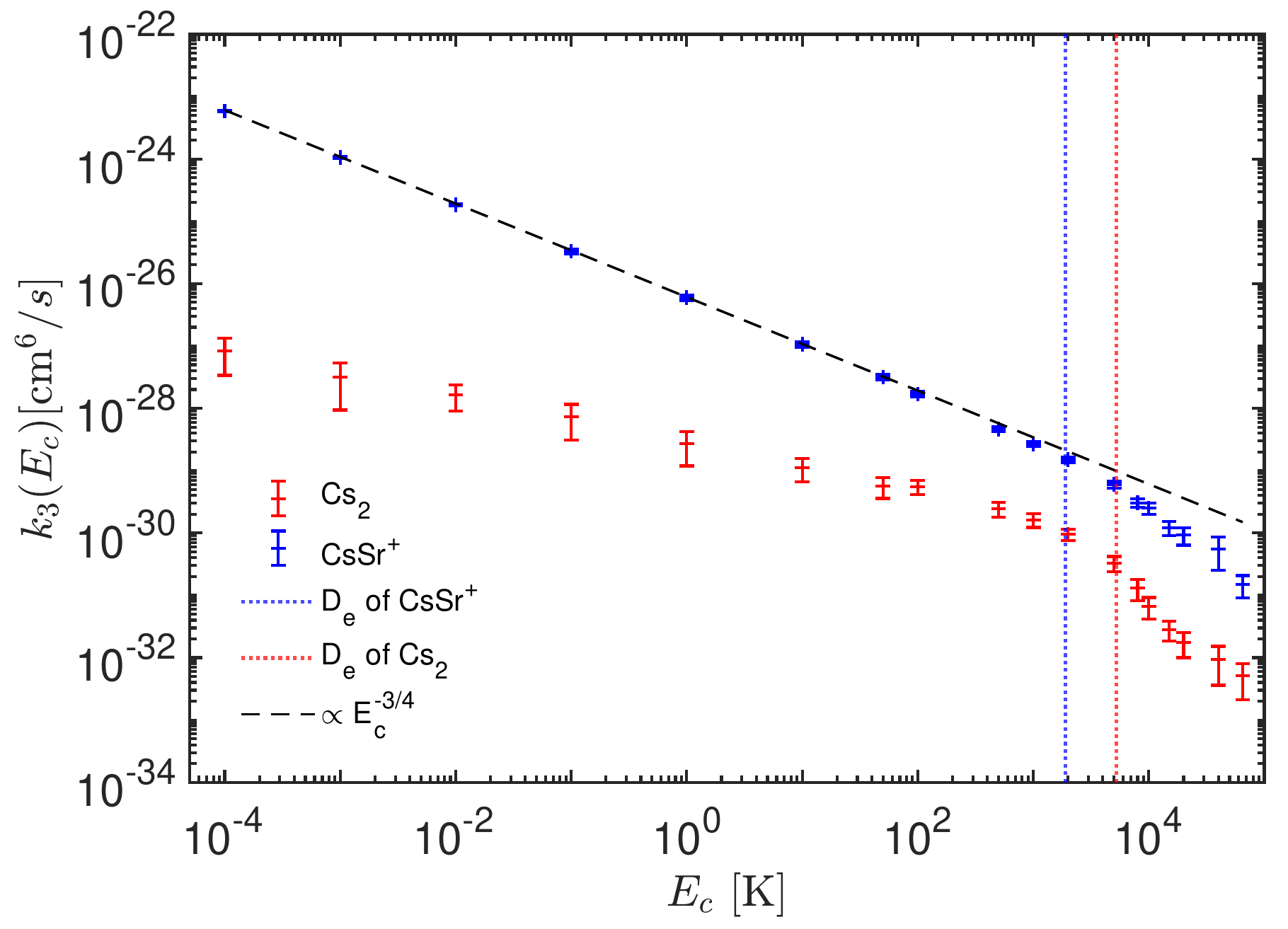}
			\caption{\label{fig:CsSr} Three-body recombination rates $k_3(E_c)$ for the Cs+Cs+Sr$^+$ reaction. Error-bars are associated with the error in \cref{eq:opacity}. The black dashed line indicates the power-law given in \cref{eq:k3Lang}. The blue and red vertical dashed lines indicate the dissociation energies of CsSr$^+$ and Cs$_2$, respectively.}
		\end{center}
	\end{figure}

	The first system under consideration is Cs+Cs+Sr$^+$, in which we assume that the Cs$_2$ is characterized via the $X^1\Sigma^+_g$ potential with $C_6^{\mathrm{Cs}_2} \approx 6.64\times10^3$~a.u. and $C_{12}^{\mathrm{Cs}_2} \approx 6.63\times 10^8$~a.u.(see Ref.~[\onlinecite{Weickenmeier1985}]). For the ion-atom potential we take $C_4^\mathrm{CsSr^+} \approx 200$~a.u. and $C_8^\mathrm{CsSr^+} \approx 1.67\times 10^6$~a.u., corresponding to the $A^1\Sigma^+$ potential for CsSr$^+$ (see Refs.~[\onlinecite{Aymar2011,Schwerdtfeger2019}]). The obtained three-body recombination rates for CsSr$^+$ (indicated by blue color) and Cs$_2$ (red color) molecules are shown in \cref{fig:CsSr}. In this figure, looking into the CsSr$^+$ rate coefficients, we identify two regimes associated with two different power-law behaviors (linear in the log-log scale). These two regimes meet at $E_c$ equal to the dissociation energy of the CsSr$^+$ potential $D_e\approx 1888$~K ($\approx 1312$~cm$^{-1}$). Similarly, the two energy regimes can be recognized through the three-body recombination rates of Cs$_2$. However, in this case, the power-law dependence is different compared to molecular ion formation. In particular, the trend of $k_3(E_c)$ for the formation of neutral molecules changes twice, one slight change near the dissociation energy of CsSr$^+$ and a pronounced change at $E_c$ comparable to the dissociation energy of Cs$_2$, i.e., $D_e\approx 5250$~K ($\approx 3650$~cm$^{-1}$).

	At low collision energies, it is noticed in \cref{fig:CsSr} that the three-body recombination rate into Cs$_2$ is almost four orders of magnitude smaller than CsSr$^+$. Therefore, the dominant product is the molecular ion and the formation rate of the neutral molecules is negligible; thus, the power-law derived in \cref{sec3} from \cref{eq:k3Lang} very well describes the trend of $k_3(E_c)$ for CsSr$^+$ formation (see the black dashed line). However, as energy increases, the ratio between both products decreases, eventually approaching the dissociation energy of the molecular ion. At this stage, the formation of Cs$_2$ can not be neglected anymore, leading to a deviation from the derived power-law behavior via \cref{eq:k3Lang} ($\propto E_c^{-3/4}$), for $E_c$ beyond the low-energy regime. In the high-energy regime, we observe that the three-body recombination rate into neutral molecules shows a steeper dependence on the collision energy than molecular ions. This behavior is due to the difference in the short-range of the atom-ion potential $\propto r^{-8}$ and the atom-atom potential $\propto r^{-12}$, as explained in Ref.~\cite{Mirahmadi2021} for the formation of van der Waals molecules. 
	

	
\begin{figure}
		\centering
		\subfigure[\label{fig:RbSr}  Three-body collision Rb+Rb+Sr$^+$. The blue and red vertical dashed lines indicate the dissociation energies of RbSr$^+$ and Rb$_2$, respectively.]{\includegraphics[scale=0.45]{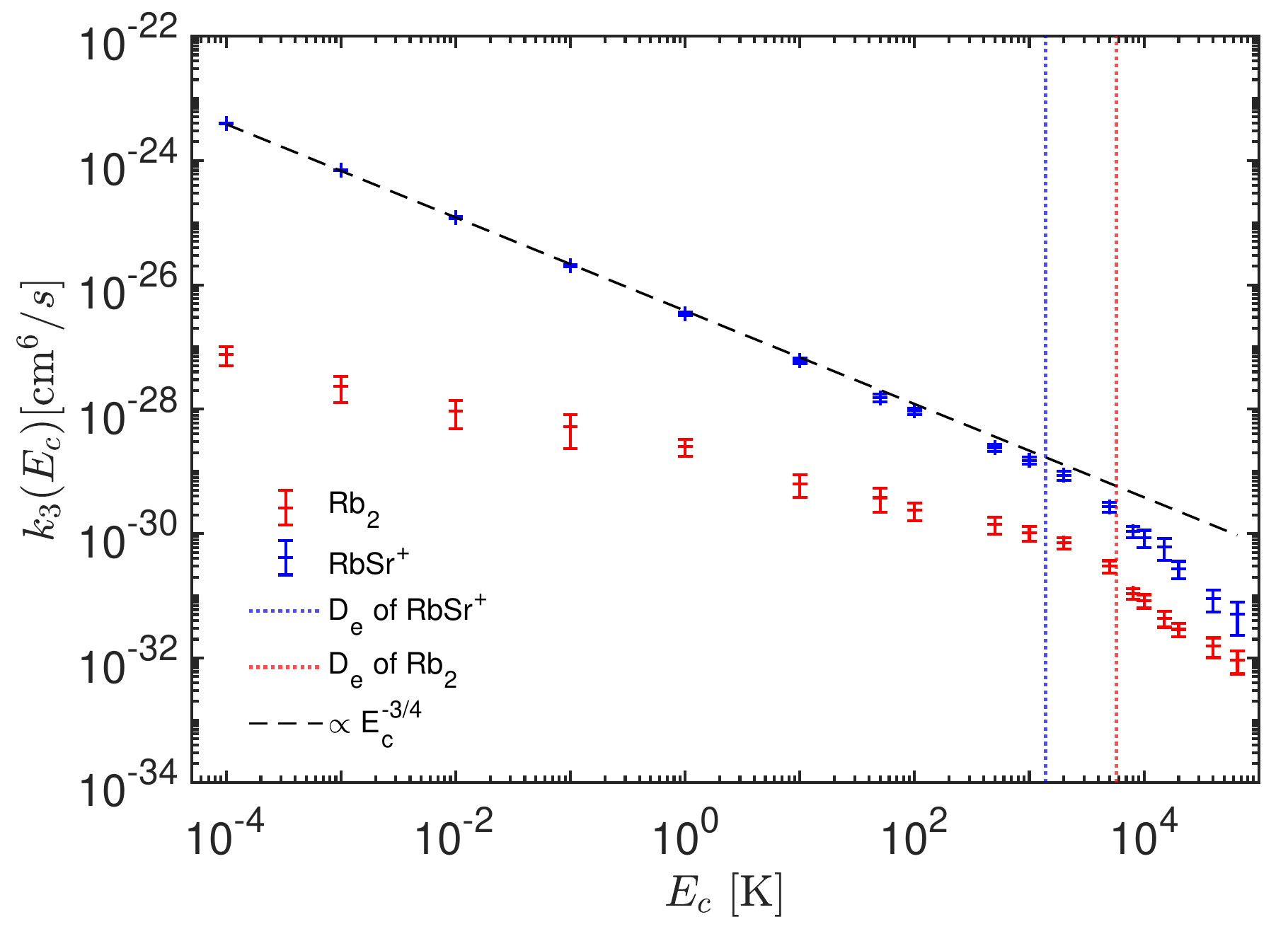}} 
		\subfigure[\label{fig:RbYb} Three-body collision Rb+Rb+Yb$^+$. The blue and red vertical dashed lines indicate the dissociation energies of RbYb$^+$ and Rb$_2$, respectively.]{\includegraphics[scale=0.45]{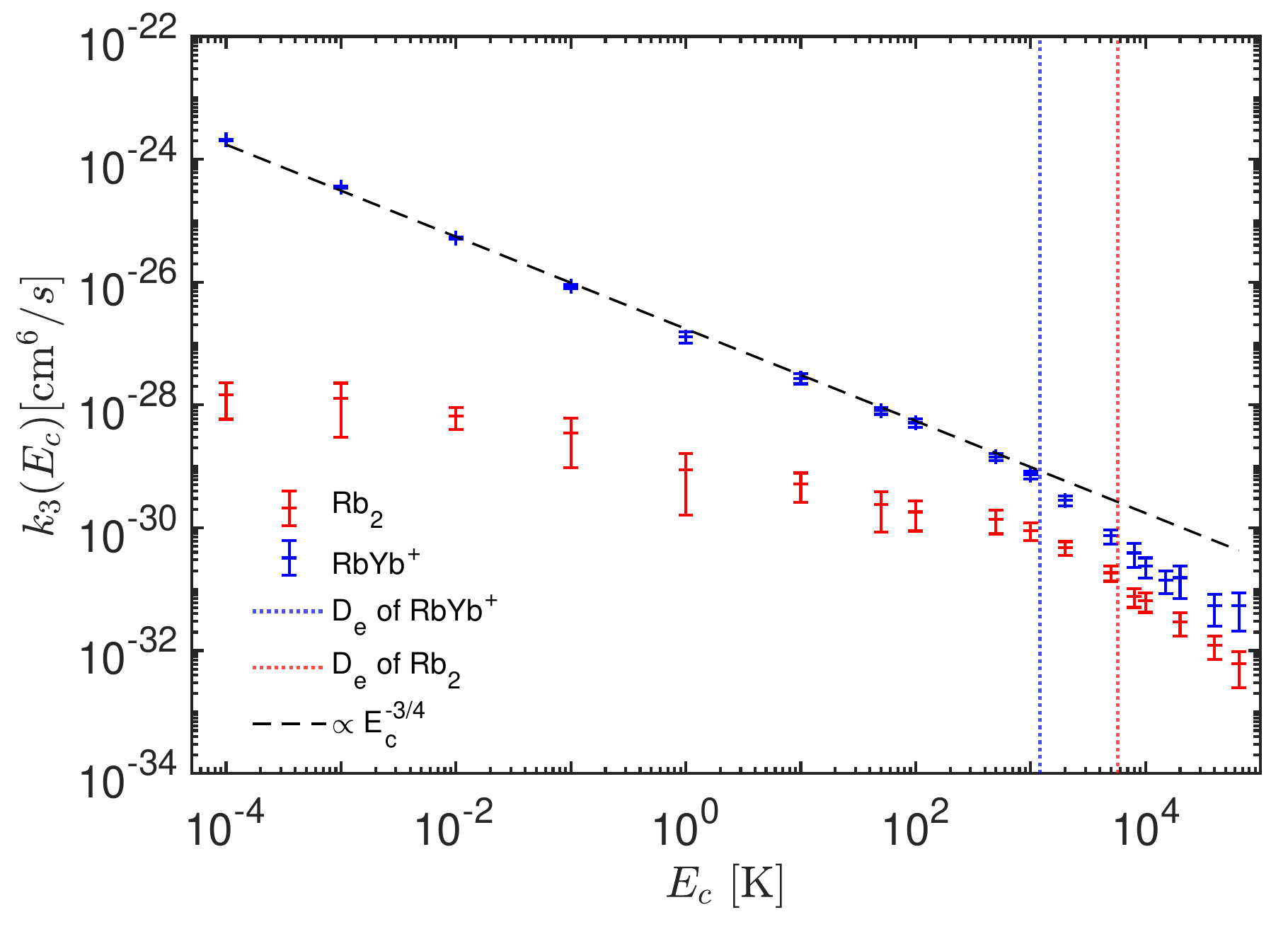}} 
		\caption{\label{fig:c4domcdf} Same as \cref{fig:CsSr} but for two systems Rb+Rb+Sr$^+$ and Rb+Rb+Yb$^+$.}
	\end{figure}

	Next, we investigate the role of the details of the short-range potential on the three-body recombination rate. In particular, we chose two systems with the same $C_6$ and $C_4$: Rb+Rb+Sr$^+$ and Rb+Rb+Yb$^+$. These systems share the same Rb$_2$ potential ($X^1\Sigma^+_g$ from Ref.~[\onlinecite{Strauss2010}]) with parameters $C_6^{\mathrm{Rb}_2} \approx 4.71\times10^3$~a.u. and $C_{12}^{\mathrm{Rb}_2} \approx 3.05\times 10^8$ in atomic units. The ion-atom potentials are taken as $A^1\Sigma^+$ with parameters $C_4^\mathrm{RbSr^+}=C_4^\mathrm{RbYb^+} \approx 160$~a.u. (from Ref.~[\onlinecite{Schwerdtfeger2019}]) and $C_8^\mathrm{RbSr^+} \approx 1.46\times 10^6$~a.u. and $C_8^\mathrm{RbYb^+} \approx 1.68\times 10^6$~a.u. (see Refs.~[\onlinecite{Aymar2011,Sayfutyarova2013}]). The results are shown in \cref{fig:RbSr,fig:RbYb}. These figures confirm the two regimes seen previously in \cref{fig:CsSr}, supporting the idea that the dissociation energy of the molecular ion is the threshold energy separating the low- from the high-energy regime.
	
	Comparing the rates illustrated in \cref{fig:RbSr,fig:RbYb}, we notice that the power-law behavior of molecular ion's recombination rates (in blue) in the high-energy limit ($E_c>D_e$) depends on the short-range properties of the two-body potentials. In contrast, the three-body recombination rates $k_3(E_c)$ in the low-energy regime ($E_c<D_e$) obey the same power-law, which confirms that low energy collisions are dominated by the long-range tail of the ion-atom potential. Note that the dissociation energy of RbSr$^+$ is $D_e\approx 1380$~K ($\approx 960$~cm$^{-1}$) and that of RbYb$^+$ is $D_e\approx 1203$~K ($\approx 836$~cm$^{-1}$). Therefore, the ratio of the products in the low energy regime is almost independent of the short-range region of the atom-atom and ion-atom two-body potentials. On the other hand, similarly to Cs+Cs+Sr$^+$ collisions, in the high-energy regime, the formation rate of neutral molecules becomes more pronounced and competes with the formation rate of AB$^+$.

	
		
	

\begin{figure}[t]
		\begin{center}
			\includegraphics[scale=0.45]{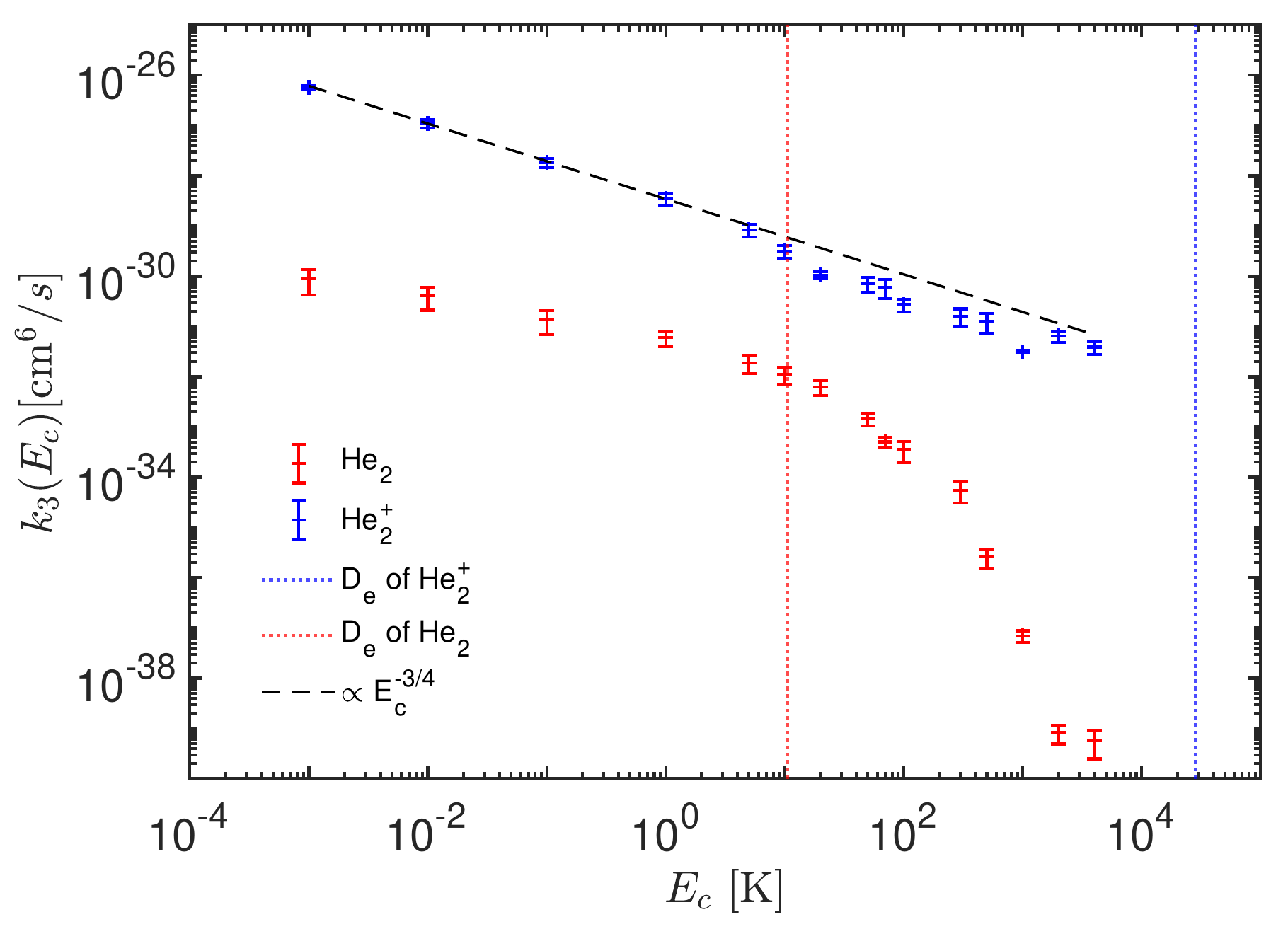}
			\caption{\label{fig:He2P} Same as \cref{fig:CsSr} but for He+He+He$^+$ three-body recombination. The blue and red vertical dashed lines indicate the dissociation energies of He$_2^+$ and He$_2$, respectively.}
		\end{center}
	\end{figure}

	To confirm the generality of the discussion above, we consider the He+He+He$^+$ three-body recombination reaction, which is in the regime associated with $\beta=-4$, although for a small $C_4$ value (in the lower left part of the diagram in \cref{fig:Pdiag}). The He$_2$ potential is taken from Ref.~[\onlinecite{Aziz1995}] with $C_6^{\mathrm{He}_2}\approx 1.35$~a.u. and dissociation energy $D_e\approx 4.5\times10^5$ and the He$_2^+$ potential is from Refs.~[\onlinecite{Hulburt1941,Chang2003}]. The energy-dependent three-body recombination rate is calculated for collision energies between 1~mK and 10$^4$~K and is displayed in \cref{fig:He2P}. 

	Unlike previous systems, in this case, the $D_e$ of the molecule (He$_2$) is smaller than the $D_e$ of the molecular ion (He$_2^+$). We notice an abrupt drop in the He$_2$ formation rate for collision energies larger than the dissociation energy of the molecule. On the contrary, the molecular ion formation rate follows the prescribed $E_c^{-3/4}$ (black dashed line) threshold law. However, we notice some deviations for collision energies larger than the dissociation energy of the molecule. This effect is so intriguing that it will be the subject of future work.
	
		
	Finally, based on our results, it is confirmed that the formation rate of molecular ions in the low-energy regime is dominated by the long-range tail of the potentials and shows the same trend ( $\propto E_c^{-3/4}$ with $\beta=-4$), independent of the A and B$^+$ species under consideration. However, this is not true for the reactions with collision energies beyond this regime, and hence, it is necessary to consider both reactions  A+A+B$^+ \rightarrow$ A+AB$^+$ and A+A+B$^+ \rightarrow$ A$_2$+B$^+$ in this regime. In particular, from the He+He+He$^+$ system, we conclude that the dissociation energy of AB$^+$ marks the limit of the low-energy regime. Thus, explaining why the threshold law is still fulfilled in noble gas ions in their parent gases at 300~K~\cite{Perez-Rios2015,Perez-Rios2018}.

	\section{Conclusions and prospects}\label{sec:conclusion}
	
This work presents a study on ion-atom-atom three-body recombination using classical trajectory calculations in hyperspherical coordinates for collision energies ranging from 100~$\mu$K to 10$^5$~K. First, we have studied the parameter space extensively for long-range atom-atom and ion-atom potentials combinations to find the behavior of the three-body long-range potential characterized by the $\beta$ parameter. $\beta$ can take any value between -4 (atom-ion dominated) and -6 (atom-atom dominated). As a result, it is possible to find three-body long-range potentials that depend on the interparticle distance differently than the underlying pairwise interaction potential ($\beta=-5$). Moreover, the value of $\beta$ relates to the production of molecules versus molecular ions. In particular, for $\beta=-4$, the production of molecular ions governs the reaction dynamics. In contrast, for $\beta=-5$ and $\beta=-6$ we find a comparable molecular formation rate between molecules and molecular ions and larger formation of molecules than molecular ions, respectively.

	Next, we have studied four distinct ion-atom-atom systems, namely,  Cs + Cs + Sr$^+$, Rb + Rb + Sr$^+$,  Rb + Rb + Yb$^+$ and He + He + He$^+$. Considering our results, we conclude the following: 
	
	\begin{itemize}
	
	    \item Every charged-neutral-neutral, A+A+B$^+$, three-body recombination reaction shows a low and high energy regime.

	    \item The low collision energy regime is described by the $\beta$ parameter, which characterizes the three-body long-range tail of the potential.

	    \item In the high-energy regime, the three-body recombination rate shows a steeper trend as a function of the collision energy than in the low-energy regime. This behavior is due to the role of short-range atom-atom and atom-ion potentials in the reaction dynamics. As a result, we observe that the reaction rate for the production of molecular ions and neutral molecules is of the same order of magnitude, in stark contrast with the low-energy regime. 
	    
	    \item The low and high energy regimes meet at collision energies comparable to the dissociation energy of the molecular ion. In other words, the dissociation energy of the main reaction product establishes the transition energy between the low and the high energy regimes. 
	    
	\end{itemize}

	
Our results refer to the probability that a given product appears as a consequence of a three-body recombination reaction. Moreover, once a neutral molecule or molecular ion appears, it can undergo dissociation or quenching processes via interactions with other particles. These effects must be included for a proper simulation of the reaction dynamics. On the other hand, at very high collision energies, many-body effects in the ion-atom-atom potential energy surface may be relevant, a topic we plan to work on shortly. Finally, our findings reveal a universal trend in ion-atom-atom three-body recombination relevant in many fields: cold chemistry, chemical physics, astrochemistry and plasma physics.

	\begin{acknowledgments}
		Authors acknowledge the support of the Deutsche Forschungsgemeinschaft (DFG – German Research Foundation) under the grant number PE 3477/2 - 493725479. J. P.-R. acknowledges the support of the Simons Foundation.
	\end{acknowledgments}
	
	
	\bibliographystyle{apsrev4-2}
	\bibliography{longrange.bib}
\end{document}